\documentstyle[prl,preprint,aps]{revtex}
\tightenlines
\begin{document}
\title{Precise measurement of hyperfine intervals using 
avoided crossing of dressed states}
\author{Umakant D. Rapol and Vasant 
Natarajan\thanks{Electronic address: 
vasant@physics.iisc.ernet.in}}
\address{Department of Physics, Indian Institute of 
Science, 
Bangalore 560 012, INDIA}

\maketitle
\begin{abstract}
We demonstrate a technique for precisely measuring 
hyperfine intervals in alkali atoms. The atoms form a 
three-level $\Lambda$ system in the presence of a strong 
control laser and a weak probe laser. The dressed states 
created by the control laser show significant linewidth 
reduction. We have developed a technique for Doppler-free 
spectroscopy that enables the separation between the 
dressed states to be measured with high accuracy even in 
room-temperature atoms. The states go through an avoided 
crossing as the detuning of the control laser is changed 
from positive to negative. By studying the separation as 
a function of detuning, the center of the level-crossing 
diagram is determined with high precision, which yields 
the hyperfine interval. Using room-temperature Rb vapor, 
we obtain a precision of 44 kHz. This is a significant 
improvement over the current precision of $\sim$1 MHz.
\end{abstract}

There has been much excitement in recent times in the use 
of atomic coherences to modify the properties of light 
beams passing through an atomic vapor. For example, 
atomic coherences have been used to slow light to low 
velocities and even to stop and ``store'' light 
\cite{PFM01}. In electromagnetically induced transparency 
(EIT) experiments, an initially absorbing medium becomes 
transparent to a probe beam when a control laser is 
applied to an auxiliary transition \cite{BIH91}. The 
control laser creates atomic coherences that shift the 
absorption away from the line center. EIT techniques have 
several practical applications in probe amplification 
\cite{MEA96}, lasing without inversion \cite{ZLN95} and 
suppression of spontaneous emission 
\cite{GZM91,ZNS95,AGA96}. In many of these experiments, 
the atoms form a three-level system and the control laser 
strongly drives two of these levels. As is well known, 
the strong driving creates dressed states of the system 
\cite{COR77}. It is the coherence between the dressed 
states that is the basis for the phenomena mentioned 
above. Experimental observations of these phenomena have 
been facilitated by the advent of low-cost tunable diode 
lasers which can be used to access transitions in alkali 
atoms such as Rb and Cs. Alkali atoms have convenient 
energy levels with strong oscillator strengths which form 
almost ideal three-level systems.

In this paper, we show that such ``coherent control'' 
techniques can be adapted to make precise measurements of 
hyperfine intervals in alkali atoms. There are two 
properties of the dressed states that are important for 
this measurement. First, the coherence between the 
dressed states results in linewidth narrowing well below 
the natural linewidth \cite{RWN02}. Hence it is possible 
to measure small shifts in the location of the dressed 
state. Second, the dressed states have opposite symmetry. 
Therefore, as the detuning of the control laser is 
changed from positive to negative, the dressed states go 
through a characteristic avoided crossing. These two 
features can be combined to determine the exact center of 
the level-crossing diagram, which yields the hyperfine 
interval. In our case, we work with room-temperature Rb 
vapor. The linewidth of the hyperfine peaks is about 27 
MHz, but the technique allows us to extract the hyperfine 
interval with a precision of 44 kHz.

To understand our technique, let us first consider the 
three-level $\Lambda$-type system in Rb in some detail. 
The measurements are done on the $D_2$ line in $^{87}$Rb 
($5S_{1/2} \leftrightarrow 5P_{3/2}$ transition). As 
shown in Fig.\ 1, $^{87}$Rb has two hyperfine levels in 
the ground state with $F=1$ and 2. The excited state 
splits into four hyperfine levels, of these the $F'=2$ 
level couples to both ground levels and can be used to 
form the $\Lambda$ system. The control laser drives the 
$F=1 \leftrightarrow F'=2$ transition with Rabi frequency 
of $\Omega_R$ and detuning from resonance of $\Delta_c$. 
The weak probe laser measures absorption on the $F=2 
\leftrightarrow F'=2$ transition at a detuning $\Delta$. 
The spontaneous decay rate from the excited level is 
$\Gamma$, which is $2\pi \times 6.1$ MHz in Rb. 

The absorption of the weak probe in the presence of the 
control laser has been derived previously 
\cite{AGA96,VAR96}. As is well known, the strong control 
laser creates two dressed states due to the ac Stark 
shift \cite{COR77}. The probe absorption gets modified 
due to this and shows peaks at the location of the two 
dressed states (Autler-Townes doublet), given by 
\begin{equation}
\Delta_{\pm} = \frac{\Delta_c}{2} \pm 
\frac{1}{2}\sqrt{\Delta_c^2 + \Omega_R^2}. 
\end{equation}
Here $\Delta_{+}$ and $\Delta_{-}$ are the values of the 
probe detuning where the peaks occur. The corresponding 
linewidths ($\Gamma_{\pm}$) of these peaks are different 
because of the coherence between the two dressed states, 
and given by
\begin{equation}
\Gamma_{\pm} = \frac{\Gamma}{2} 
\left( 1 \mp \frac{\Delta_c}{\sqrt{\Delta_c^2 + 
\Omega_R^2}} \right) .
\end{equation}
It is clear from the above expression that, if $\Delta_c 
= 0$, the two peaks are symmetric and have identical 
linewidths of $\Gamma /2$. However, for any non-zero 
detuning, the peaks have asymmetric linewidths. The first 
peak has larger linewidth while the second peak has 
smaller linewidth by precisely the same factor, in such a 
way that the sum of the two linewidths is equal to the 
unperturbed linewidth, $\Gamma$.

The above analysis shows how the three-level system is 
useful in many applications. For example, probe 
absorption at line center ($\Delta=0$) is strongly 
suppressed in the presence of a resonant control laser 
because the dressed states created by the control laser 
are shifted by the Rabi frequency. This is the basis for 
EIT experiments. Similarly, it is clear from Eq.\ 2 that 
the linewidth $\Gamma_+$ of the second dressed state can 
be much below the natural linewidth when the 
control-laser detuning is large. The basic idea for our 
experiment is also contained in the above analysis. From 
Eq.\ 1, the separation between the dressed states is 
$\sqrt{\Delta_c^2 + \Omega_R^2}$. As the detuning of the 
control laser is varied from negative to positive, the 
states exhibit an avoided crossing, so that the 
separation at zero detuning is not zero but equal to the 
Rabi frequency. Thus, by studying the variation of the 
separation with detuning, it is possible to determine the 
location of the level crossing precisely.

However, observing the Autler-Townes doublet in 
room-temperature vapor is complicated by effects due to 
Doppler broadening. The above expressions are valid for a 
stationary atom; in room-temperature vapor they have to 
be corrected for the thermal velocity distribution of the 
atoms. One consequence of this is that the linewidth 
appearing in Eq.\ 2 is not the natural linewidth but the 
Doppler width, which is 560 MHz for room-temperature Rb 
atoms. We have solved the problem of Doppler broadening 
in the following manner. A part of the probe laser is 
split off as a ``pump'' beam and sent through the vapor 
cell so that it is counter-propagating with respect to 
the probe and control beams. The intensity of the pump 
beam is chosen to be about 4 times higher than the probe. 
In this configuration, the zero-velocity group of atoms 
preferentially absorbs from the pump and the probe gets 
transmitted. This is a standard technique used in 
Doppler-free saturated-absorption spectroscopy 
\cite{DEM82} which we have adapted to the three-level 
case.

The experimental schematic for the hyperfine measurements 
in Rb is shown in Fig.\ 2. The probe and control beams 
are obtained from two frequency-stabilized diode laser 
systems operating near the $D_2$ line in $^{87}$Rb. The 
linewidth of the lasers after stabilization has been 
measured to be below 1 MHz. The two beams co-propagate 
through a room-temperature vapor cell containing Rb. The 
absorption through the cell is about 25\%. The 
counter-propagating pump beam discussed above is generated from 
the probe laser using a beamsplitter. To set the 
frequency of the control laser, a part of the control 
beam is tapped off for saturated-absorption spectroscopy 
in a vapor cell. It can be potentially locked to any of 
the hyperfine peaks or crossover resonances in the 
spectrum. We lock it to the $F=1 \leftrightarrow 
F'=(1,2)$ crossover resonance. The remaining frequency 
offset (of about 78 MHz) to bring it close to the $F=1 
\leftrightarrow F'=2$ is obtained using an acousto-optic 
modulator (AOM). The frequency shift in the AOM is varied 
to get different detunings and the value of the AOM 
frequency at which the level-crossing occurs gives the 
hyperfine interval between the $F=(1,2)$ crossover 
resonance and the $F'=2$ level. Note that the only 
absolute frequency entering the measurement is the 
frequency of the AOM, all other frequencies such as the 
separation of the dressed states or the Rabi frequency of 
the control laser are only important in their relative 
values.

The transmission of the probe laser as it is scanned 
across the $F=2 \leftrightarrow F'=2$ transition is shown 
in Fig.\ 3. In the first trace, the control laser is 
turned off. The pump-probe configuration results in the 
appearance of a Doppler-free peak as for standard 
saturated-absorption spectroscopy. Ideally, the linewidth 
of the peak should be the natural linewidth of 6.1 MHz. 
The linewidth obtained in our case is about 27 MHz. There 
are many reasons for this broadening, the main two being 
misalignment between the counter-propagating beams, and 
power broadening due to the pump beam. However, as we 
will see below, the large linewidth does not seriously 
affect the measurement.

In the presence of the control laser, the $F'=2$ peak 
splits into two peaks exactly as predicted by the above 
analysis. This is shown in the middle trace of Fig.\ 3. 
The two peaks have asymmetric linewidths because of the 
non-zero detuning of the control laser, with the peak on 
the right having smaller linewidth. The detuning for this 
spectrum is +13.5 MHz. In the lowest trace of Fig.\ 3, we 
show the same spectrum when the detuning has been changed 
to $-13.5$ MHz. Notice that now it is the peak on the 
left that has the smaller linewidth, showing that the 
symmetry of the dressed states changes as the detuning 
changes from positive to negative. Furthermore, the sum 
of the linewidths of the two peaks in both cases is equal 
to the unperturbed linewidth, as predicted by Eq.\ 2. 
However, as mentioned before, the unperturbed linewidth 
is 27 MHz and not the natural linewidth of 6.1 MHz.

The variation of the separation of the Autler-Townes 
doublet as a function of AOM frequency is shown in Fig.\ 
4. Since the frequency of the control laser is first 
shifted using the AOM, and then locked to the $F=1 
\leftrightarrow F'=(1,2)$ crossover resonance, the AOM 
frequency at which the separation reaches a minimum 
yields the hyperfine interval to the $F'=2$ level. The 
solid line in Fig.\ 4 is the expected variation in 
separation as a function of control-laser detuning from 
Eq.\ 1. The best fit yields a value of {\bf 78.433(44)} 
MHz for this hyperfine interval. Note that the absolute 
values of the control-laser Rabi frequency and the 
separation of the peaks are not important. Any scaling 
error in obtaining these numbers from the measured 
experimental parameters will not affect the determination 
of the minimum. The only absolute number of consequence 
is the AOM frequency and this is measured very precisely 
using a frequency counter.

The accuracy of 44 kHz we obtain is significantly better 
than the accuracy with which hyperfine intervals in 
alkali atoms are currently known \cite{AIV77}. The values 
in Ref.\ \cite{AIV77} are obtained by fitting to all the 
available data on alkali atoms with measurements from a 
range of techniques such as optical double resonance, 
cascade radiofrequency and level crossing. The typical 
error in Ref.\ \cite{AIV77} for hyperfine intervals in 
the $5P_{3/2}$ state of Rb is about 1 MHz. There is a 
more recent measurement in Rb \cite{BGR91} using 
stabilized diode lasers similar to the lasers we have 
used in this work. Absolute frequencies of the different 
hyperfine levels are obtained using a Fabry-Perot 
interferometer and a stabilized HeNe laser as the 
frequency reference. The accuracy quoted in Ref.\ 
\cite{BGR91} is 0.4 MHz. The frequency stability of the 
diode lasers is a limitation in obtaining higher accuracy 
with such an interferometric technique. 

The most precise measurements in $^{87}$Rb to date have 
been reported in Ref.\ \cite{YSJ96} with a quoted error 
of only 4--9 kHz. The extremely small error in this work 
has been achieved using optical heterodyning of two 
ultra-stable tunable Ti-sapphire lasers locked to 
different hyperfine transitions. Several experimental 
advancements were used to bring down errors to this 
level. The linewidth of the hyperfine transitions was 
reduced to 7 MHz in the vapor cell using magnetic 
shielding and careful control of pump and probe 
intensities. The cells themselves were specially 
constructed to ensure ultra-high purity. The lasers were 
locked with 3 kHz precision to the line center using 
third-harmonic lock-in detection at the modulation 
frequency. Ultra-stable frequency standards were used to 
achieve stability of AOMs. By contrast, we have used 
standard locking of a low-cost diode laser to the 
hyperfine transition. The transition is broadened to a 
linewidth of 27 MHz in the vapor cell. We have therefore 
not really pushed the limits of precision for our 
technique. It is conceivable that, by using some of the 
advanced techniques mentioned in Ref.\ \cite{YSJ96}, we 
can reduce our errors by a factor of 5 to 10. However, 
even at the level of 44 kHz, the technique is useful 
since hyperfine intervals in the other isotope of Rb, 
$^{85}$Rb, or in other alkali atoms are known only with 
$\sim$MHz precision.

In conclusion, we have demonstrated a new technique for 
measuring hyperfine intervals in alkali atoms. The atoms 
form a three-level $\Lambda$ system with an excited level 
coupled to a ground level by a strong control laser, and 
the same excited level coupled to a different ground 
level by a weak probe laser. The control laser creates 
dressed states that show significant linewidth reduction 
when the control laser is detuned from resonance. In 
addition, the states go through an avoided crossing as 
the sign of the detuning is changed from positive to 
negative. Even though we work with room-temperature 
vapor, we have adapted a technique of using a 
counter-propagating pump beam that allows us to overcome the 
Doppler effect and measure the separation of the dressed 
states very accurately. By studying the separation as a 
function of detuning, we determine the line center of the 
level-crossing diagram with high precision. We work with 
room-temperature Rb atoms, where we obtain a precision of 
44 kHz. This is already a useful level of precision since 
typical accuracy for hyperfine intervals in alkali atoms 
is of order 1 MHz. However, we think that with 
foreseeable improvements, the error can be brought down 
to the few kHz level. The technique is easily extended to 
other alkali atoms such as Li, K, and Cs, where the 
transitions are accessible with low-cost tunable diode 
lasers.

The authors thank Anusha Krishna for help with the 
experiments. This work was supported by a research grant 
from the Department of Science and Technology, Government 
of India.

\begin{figure}
\caption{
Three-level $\Lambda$ system in $^{87}$Rb. The $F'=2$ 
hyperfine level in the excited state is coupled to two 
ground hyperfine levels: to $F=1$ by the control laser 
and to $F=2$ by the probe laser. The control laser has 
detuning $\Delta_c$ and Rabi frequency $\Omega_R$. The 
probe laser has detuning $\Delta$. The spontaneous decay 
rate from the excited level is $\Gamma$.
}
\label{fig1}
\end{figure}

\begin{figure}
\caption{
Schematic of the experiment. The probe and control beams 
are derived from frequency-stabilized, tunable diode 
laser systems. The power in each beam is set using a 
half-wave plate ($\lambda/2$) and a polarizing 
beamsplitter (PBS). The two beams have orthogonal 
polarizations and are mixed in a PBS, co-propagate 
through the Rb cell, and are separated again using a PBS. 
The probe beam is detected on a silicon photodetector 
(PD). A part of the probe beam is split using a 
beamsplitter (BS) and sent through the cell so that it 
counter-propagates with respect to the probe and control 
beams. The angle between the counter-propagating beams 
has been exaggerated for clarity; in reality it is close 
to 0 ensuring good overlap of the beams in the cell. The 
frequency of the control laser is set by first shifting 
it using an acousto-optic modulator (AOM) and then 
locking to a hyperfine peak in the saturated-absorption 
spectrum. All unwanted beams terminate on beam dumps 
(BD).
}
\label{fig2}
\end{figure}

\begin{figure}
\caption{
Autler-Townes doublet. The figure shows probe 
transmission spectra as a function of probe detuning. The 
top trace is with the control laser off and shows a 
standard saturated-absorption spectrum of the $F'=2$ 
peak. The linewidth of the peak is 27 MHz. In the middle 
trace, the control laser is turned on at a detuning of 
+13.5 MHz. The peak splits into an Autler-Townes doublet, 
with linewidths of 18 MHz and 9 MHz respectively. The 
bottom trace is for a control-laser detuning of -13.5 
MHz. The symmetry of the peaks has changed, and the 
linewidths are now 9 MHz and 17 MHz respectively. Slow 
varying Doppler profiles have been removed from all 
traces.
}
\label{fig3}
\end{figure}

\begin{figure}
\caption{
Separation vs. AOM frequency. The figure shows the 
variation in the separation of the Autler-Townes doublet 
peaks as a function of AOM frequency. The minimum 
separation occurs when the control-laser detuning is 
zero. The solid line is a fit to the expected variation 
from Eq.\ 1 in the text, which yields a value of 
78.433(44) MHz for the interval between the $F'=2$ level 
and the $F'=(1,2)$ crossover resonance.
}
\label{fig4}
\end{figure}

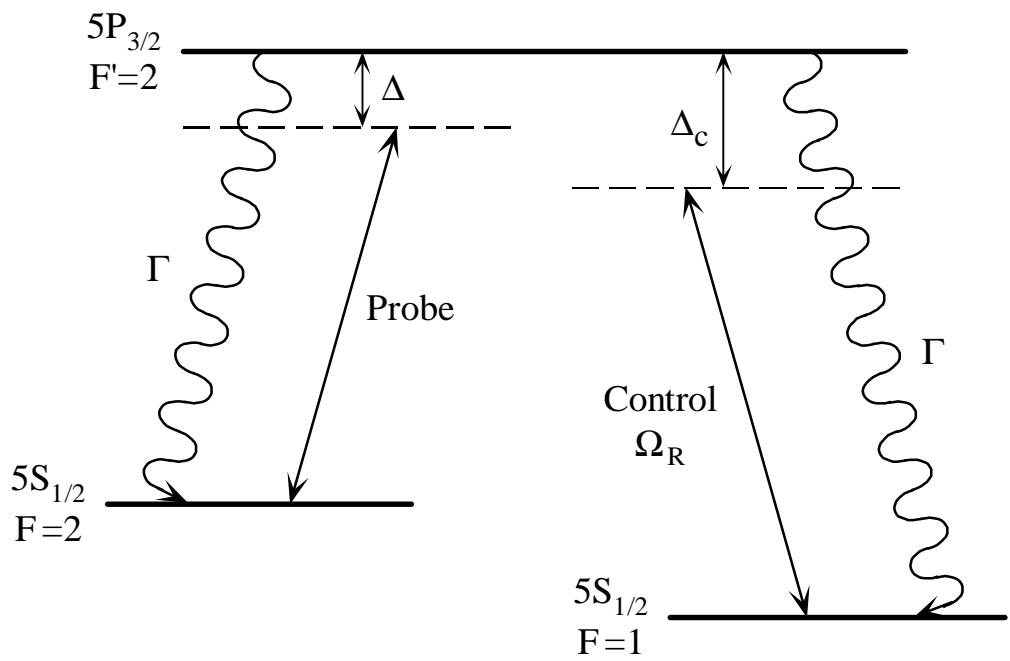
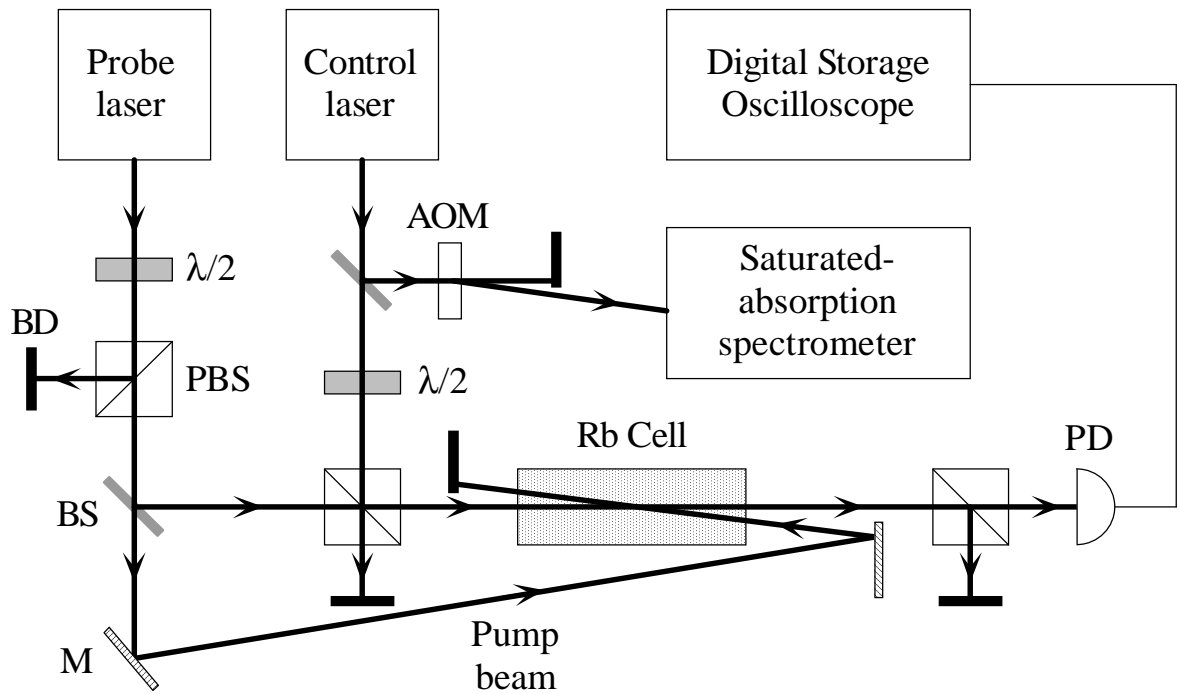
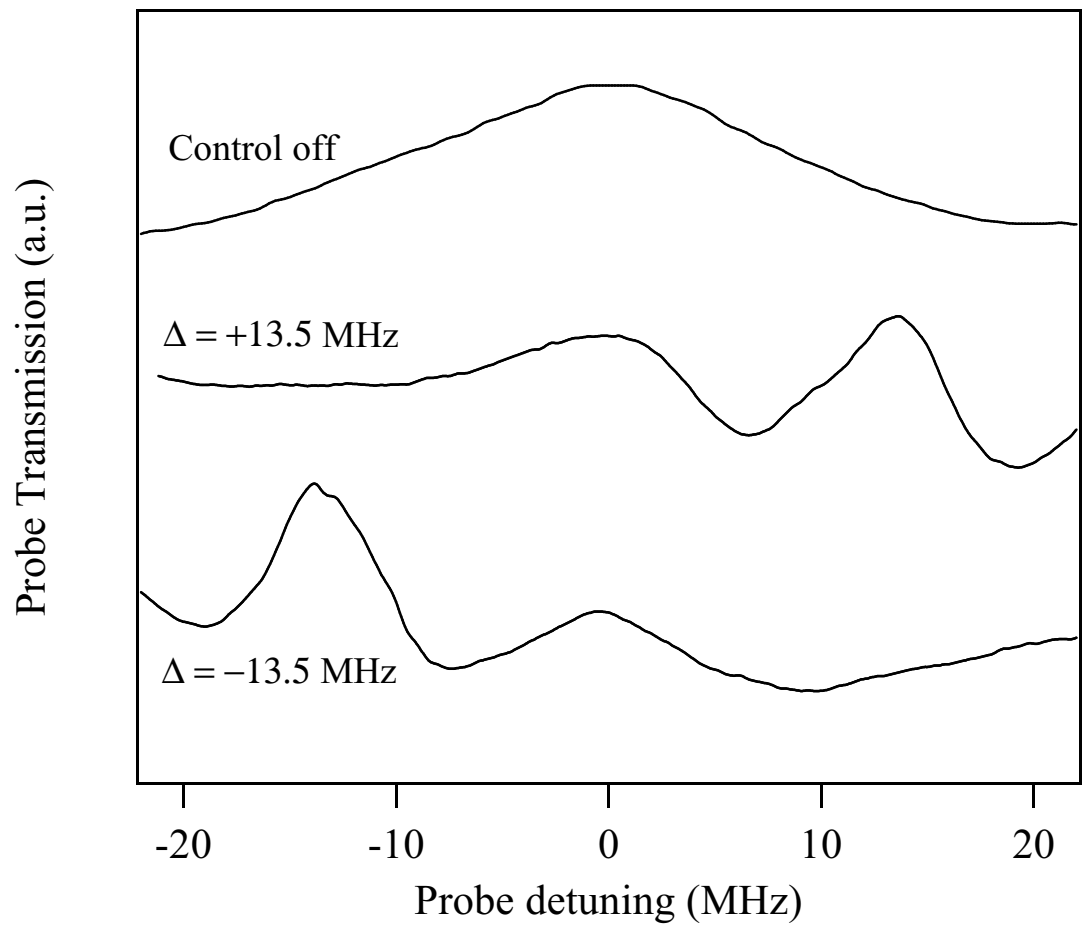
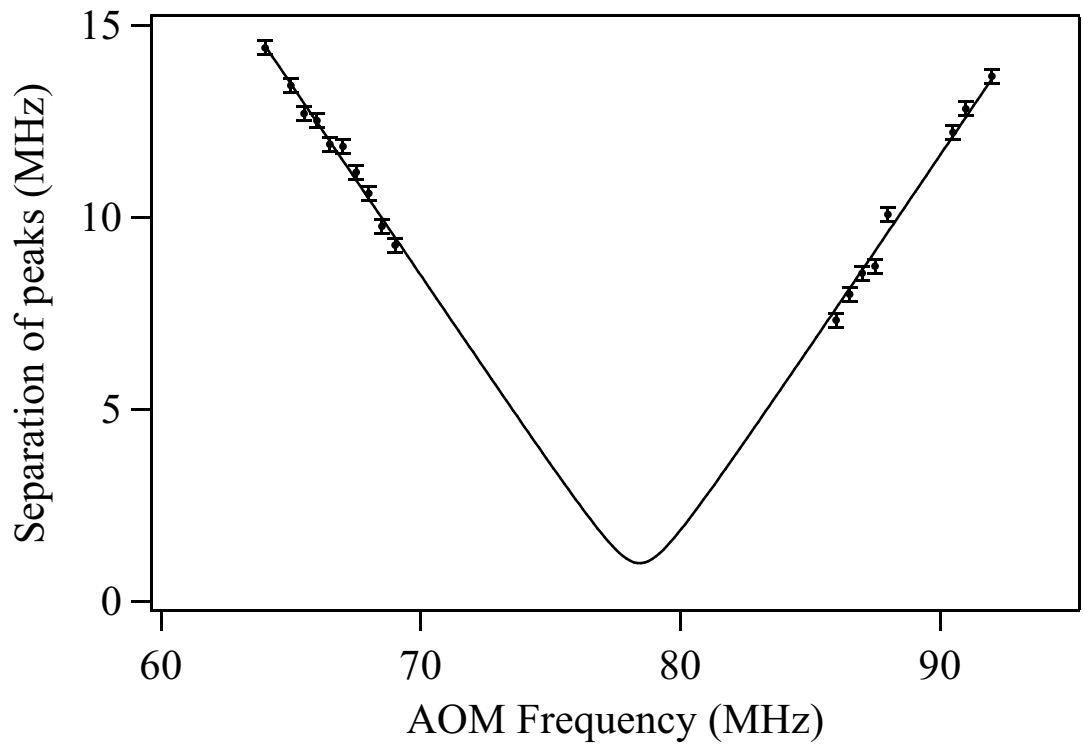
\end{document}